\documentclass[twocolumn]{aastex63}
\usepackage{graphicx}
\usepackage{color}
\usepackage{amsmath,amssymb}
\usepackage{gensymb}
\usepackage{times}
\usepackage{nicefrac}
\usepackage{xcolor}
\usepackage{url}

\usepackage[utf8]{inputenc}

\begin{document}
\title{Stellar Mergers or Truly Young?\\
\vspace{0.1in}
Intermediate-Age Stars on Highly-Radial Orbits in the Milky Way's Stellar Halo}

\author[0000-0003-1856-2151]{Danny Horta}
\affiliation{Center for Computational Astrophysics, Flatiron Institute, 162 5th Ave., New York, NY 10010, USA\\}

\newcommand{\affcolumbia}{
    Department of Astronomy, Columbia University,
    550 West 120th Street, New York, NY 10027, USA
}

\newcommand{\affamnh}{
Astrophysics Department, American Museum of Natural
History, 200 Central Park West, New York, 10024, NY, USA
}

\newcommand{\affanu}{
Research School of Astronomy $\&$ Astrophysics, Australian National University, Canberra ACT 2611, Australia
}

\author[0000-0003-4769-3273]{Yuxi (Lucy) Lu}
\affiliation{\affamnh}
\affiliation{\affcolumbia}

\author[0000-0001-5082-6693]{Melissa K. Ness}
\affiliation{\affcolumbia}
\affiliation{\affanu}

\author[0000-0002-8495-8659]{Mariangela Lisanti}
\affiliation{Center for Computational Astrophysics, Flatiron Institute, 162 5th Ave., New York, NY 10010, USA\\}
\affiliation{Department of Physics, Princeton University, Princeton, NJ 08544, USA\\}

\author[0000-0003-0872-7098]{Adrian M. Price-Whelan}
\affiliation{Center for Computational Astrophysics, Flatiron Institute, 162 5th Ave., New York, NY 10010, USA\\}

\correspondingauthor{Danny Horta}
\email{dhortadarrington@flatironinstitute.org}

\begin{abstract}
Reconstructing the mass assembly history of the Milky Way relies on obtaining detailed measurements of the properties of many stars in the Galaxy, especially in the stellar halo. One of the most constraining quantities is stellar age, as it can shed light on the accretion time and quenching of star formation in merging satellites. However, obtaining reliable age estimates for large samples of halo stars is difficult. We report published ages of 120 subgiant 
halo stars with highly-radial orbits that likely belong to the debris of the \textsl{Gaia-Enceladus/Sausage}~(GES) galaxy.
The majority of these halo stars are old, with an age distribution characterized by a median of 11.6~Gyr and 16$^{\rm th}$(84$^{\rm th}$) percentile of 10.5~(12.7)~Gyr.  However, the distribution is skewed, with a tail of younger stars that span ages down to $\sim6$--$9$ Gyr. All highly-radial halo stars have chemical and kinematic/orbital quantities that associate them with the GES debris.  Initial results suggest that these intermediate-age stars are not a product of mass transfer and/or stellar mergers, which can bias their age determination low. If this conclusion is upheld by upcoming spectro-photometric studies, then the presence of these stars will pose an important challenge for constraining the properties of the GES merger and the accretion history of the Galaxy.

\end{abstract}

\section{Introduction}

In addition to a star's chemical and kinematic/orbital properties, its age also serves as a useful quantity for inferring its origin. For large samples, stellar ages are crucial for understanding stellar and galactic evolution (\citealp[e.g.,][]{Edvardsson1993,Lachaume1999,Nordstrom2004,Martig2016,Ness2016,Sanders2023}). However, obtaining stellar age measurements is a non-trivial task.

Traditional recipes typically rely on using colour-magnitude diagrams~(CMD) and fitting theoretical stellar evolutionary
sequences (i.e., stellar isochrones) to stars to infer their age \citep[][]{Soderblom2010}. This technique is robust for main-sequence turn off (MSTO) and subgiant stars that extend to the oldest possible ages ($\sim14$ Gyr, \citealp[e.g.,][]{Xiang2017}). This is because there is substantial discriminating power of ischrones for measuring stellar age at this evolutionary state. Moreover, the chemical compositions of stellar atmospheres determined from spectra of subgiant stars reflect the birth gas from which they formed \citep[][]{Freeman2002}. This makes subgiants some of the best chemical-chronological-dynamical tracers \citep[][]{Xiang2022}. 

\begin{figure*}
    \centering
    \includegraphics[width=0.9\textwidth]{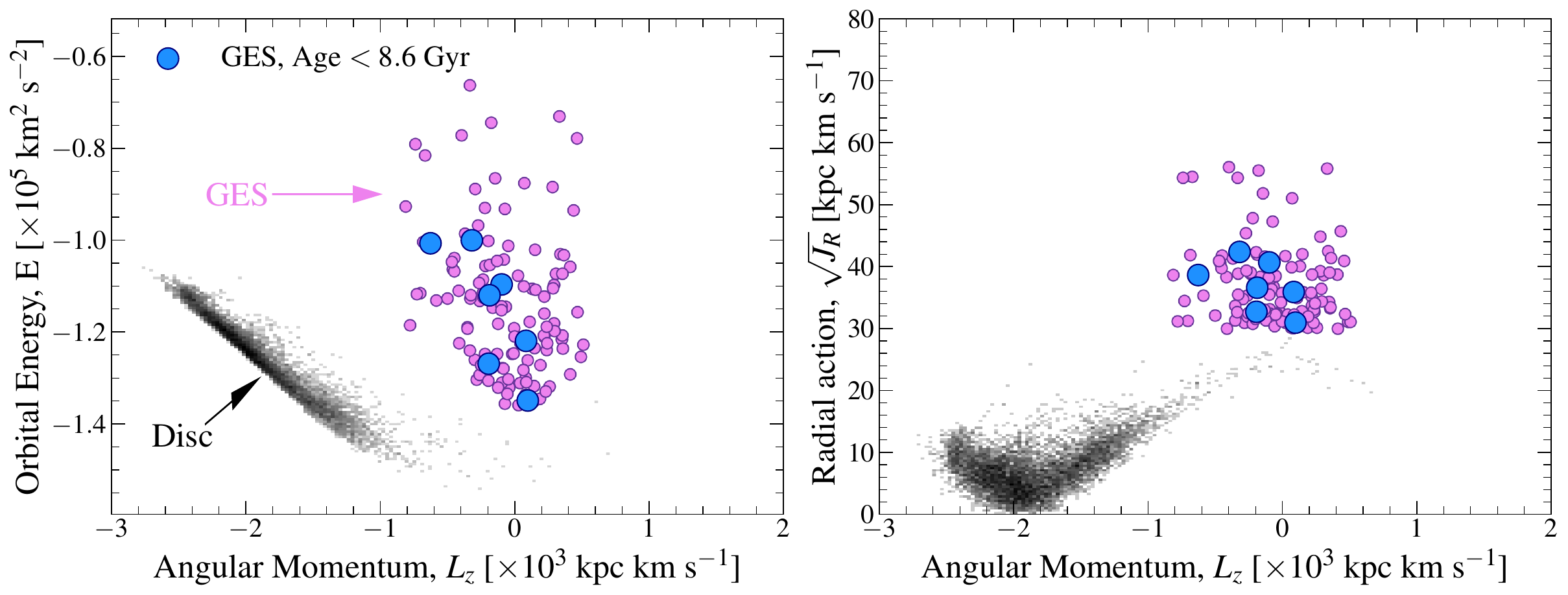}
    \caption{Dynamical selection of the GES sample. \textit{Left}:~The GES sample~(pink) and parent sample~(black) in the orbital energy versus azimuthal angular momentum plane. Shown as blue are the intermediate-age GES stars (see Section~\ref{sec_results}). GES stars sit on high-energy and low-$L_z$ orbits. Here, prograde populations have negative $L_{z}$. \textit{Right}:~Same as the left panel, but for the radial action versus $L_z$ plane, which we use to select the GES sample.}
    \label{fig:selection}
\end{figure*}

For the Milky Way, the time at which a star is born is particularly useful for retelling the mass assembly history of the Galaxy because different Galactic populations have different stellar age distributions (\citealp[e.g.,][]{Gallart2019,Spitoni2019,Xiang2022}). For example, old stars (i.e., Age~$\gtrsim10$~Gyr) tend to be associated with the Galactic stellar halo and/or thick (high-$\alpha$) disc. Conversely, younger stars (Age~$\lesssim6$~Gyr) are typically associated with the thin (low-$\alpha$) disc and are predominantly formed \textit{in situ}. Moreover, the ages of stars born \textit{ex situ} help constrain the time in which accretion events occurred in the Galaxy \citep[][]{Montalban2021}. They also help constrain the time in which these accreted systems quenched star formation \citep[][]{Bonaca2020}, as the youngest star should be the latest one formed. Thus, stellar ages are crucial for unveiling the accretion history of the Milky Way and for inferring the properties of accreted satellites.

Until recently, obtaining large samples of stellar parameter and age information for subgiant stars across the Galaxy has been difficult due to the short lifetime of their evolutionary phase. However, the advent of large-scale stellar surveys (e.g., \textsl{APOGEE}: \citealp{Majewski2017}, \textsl{LAMOST}: \citealp{Zhao2012}, \textsl{GALAH}: \citealp{Freeman2012}) and the \textsl{Gaia} mission~\citep[]{Gaia2016,Gaia2022} has supplied a wealth of data for millions of Milky Way stars, making it possible to determine reliable ages, element abundances, and orbits for large samples of subgiants. 

In this paper, we use the publicly available catalogue of stellar ages for subgiant stars compiled by \citet{Xiang2022} (hereafter, XR22) using the \textsl{LAMOST} DR7 and \textsl{Gaia}~DR3 data to examine the age distribution of stars in the Galaxy on highly-radial orbits that likely belong to the remnant of the \textsl{Gaia}-Enceladus/Sausage (GES) accretion event (\citealp[][]{Belokurov2018,Haywood2018,Helmi2018,Mackereth2019}). We seek to investigate $i$) what is the overall distribution of ages in the GES debris, and $ii$) how old are the youngest stars in GES, with the aim of constraining the properties of the GES system and its accretion episode within the Milky Way. 

This paper is organized as follows.  In Section~\ref{sec_data}, we present the data and define our sample. In Section~\ref{sec_results}, we present the results on the element abundance, age, and kinematic/orbital properties. We then discuss the implications of our findings in Section~\ref{sec_discussion} before concluding in Section~\ref{sec_conclusions}.

\section{Data}
\label{sec_data}
We use the public catalogue of stellar ages for subgiant stars computed by XR22 using the \textsl{LAMOST} DR7 and \textsl{Gaia} DR3 surveys. This catalogue also has the element abundances from \textsl{LAMOST} and the required 6D phase-space information to determine kinematics/orbits from \textsl{Gaia/LAMOST}. In total, this sample contains 247,104 subgiant stars. 

As we are focusing on examining the ages of stars in the GES, which is primarily a metal-poor population in the Galactic stellar halo, we follow the advice from XR22 to obtain reliable age measurements for metal-poor stars and enforce the following selection criteria:

\begin{description}
    \item[High signal-to-noise spectra] \hfill \\ 
        \textsl{LAMOST} spectral S/N $> 80$,
    \item[Subgiant stars] \hfill \\ 
        \textsl{LAMOST}-determined atmospheric parameters, effective temperature and surface gravity $4000<T_{\mathrm{eff}}<7000$ K and $2 < \log(g) < 5$,
    \item[Precise age measurements] \hfill \\ 
        $\sigma_{\mathrm{age}}/\mathrm{age}<0.1$,
    \item[High-quality astrometry and stars not in wide binaries] \hfill \\ 
        \textsl{RUWE} $<1.2$,
    \item[No low-$\alpha$ disc stars] \hfill \\ 
            $\mathrm{[Fe/H]}<-0.5$.
\end{description}
These quality cuts yield a sample of 11,336 stars.

\subsection{Sample of Gaia-Enceladus/Sausage stars}

\begin{figure*}
    \centering
    \includegraphics[width=0.9\textwidth]{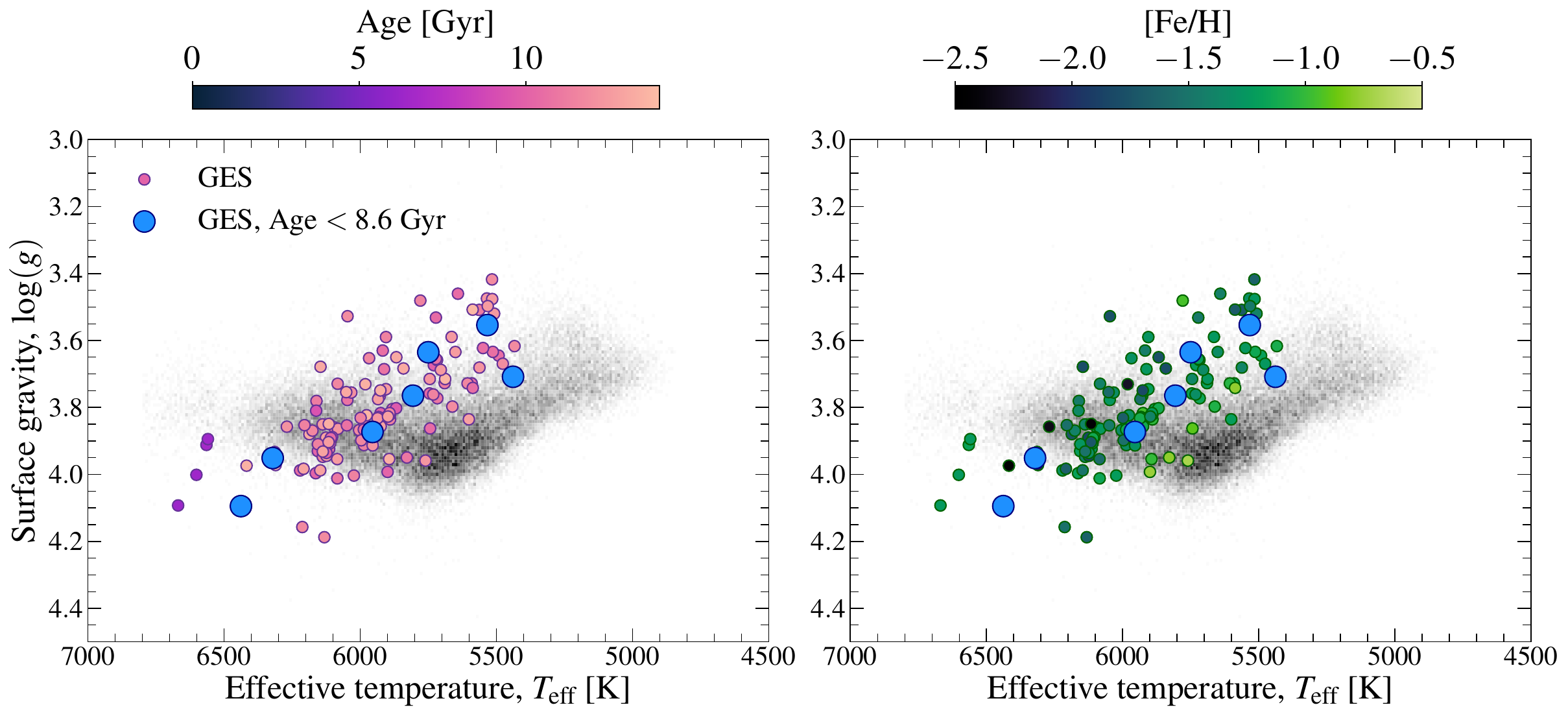}
    \caption{The GES sample~(colour) and parent sample~(black) in the Kiel diagram, where the GES sample is colour-coded by stellar age~(left) and [Fe/H] abundance~(right). As blue markers we also show the intermediate-age GES stars (see Section~\ref{sec_results}). There is a subset of hot young stars ($T_{\mathrm{eff}}>6500$ K), which we remove from the analysis (see text for details).}
    \label{fig:kiel}
\end{figure*}

To pick out stars on highly-radial orbits that are likely part of a pure sample of the GES debris, we perform a dynamical cut following a similar criteria to \citet{Feuillet2021}. We select stars with $|L_z|<700$~kpc~km~s$^{-1}$ and $\sqrt{J_R}>30$~kpc~km~s$^{-1}$. This selection yields 124 stars with highly-radial orbits.\footnote{The $J_{r}$-$L_z$ method has been pointed out by \citet{Carrillo2024} to be the most robust method of picking out a pure sample of GES stars.} We integrate orbits using the \textsl{Gala} package \citep[][]{gala2017} assuming the \texttt{MilkyWayPotential2022} potential \citep{Pricepotential2022}, which is a four-component Milky Way mass model consisting of spherical Hernquist nucleus and bulge components, an (approximate) exponential disk component, and a spherical NFW halo component. 
This potential model is fitted to the \citet{Eilers:2019} rotation curve and a compilation of Milky Way mass enclosed measurements (see Section 2 of \citealt{Hunt2022}).
This is done by converting the 6D phase-space information\footnote{Sky positions and proper motions are obtained from \textsl{Gaia} DR3 astrometry, the distances are taken from \citet{Xiang2022}, and the radial velocities are taken from \textsl{LAMOST} DR7.} into Cartesian and cylindrical coordinates assuming a position of the Sun of $R_{\odot} = 8.275$~kpc \citep[]{Gravity2022} and $z_{\odot} = 20.8$~pc \citep[]{Bennett2019}, and a local standard of rest of ($U_{\odot}$, $V_{\odot}$, $W_{\odot}$) = $(8.4, 251.8, 8.4)$ km s$^{-1}$. Orbital actions, $J=(J_R,J_{\phi},J_z)$, are then computed using the \textsl{AGAMA} package \citep{agama2019} using the ``Staeckel Fudge'' approximation \citep{Binney2012}. 

The GES sample is shown in pink/blue in the orbital energy, $E$, and radial action, $J_R$, as a function of azimuthal angular momentum, $L_z$, planes in Fig.~\ref{fig:selection}. Here, the parent sample is shown as a black 2D histogram. The GES sample occupies a locus of high-energy and radial orbits (by definition), typical of the GES debris \citep[e.g.,][]{Belokurov2018,Helmi2018,Koppelman2019,Mackereth2019,Horta2023}.

In Fig.~\ref{fig:kiel}, we also show the same data in the Kiel diagram. In the left panel, the highly-radial stars are coloured by their stellar age and in the right panel, by their [Fe/H] abundance. Overall, the highly-radial stars are hotter for each given surface gravity when compared to the parent sample, as expected for more metal-poor populations. However, there is a tail of young hot stars in the GES sample. To ensure trustworthy age measurements, we remove these four young hot-star outliers\footnote{It is possible that these hotter and younger stars could be blue stragglers.} by enforcing  $T_{\mathrm{eff}}<6500$~K. This additional cut leads to a final working sample comprised of 120 (candidate) GES stars.

Lastly, we also examined the extinctions provided by \textsl{Gaia} ($A_{G}$, $A_{\mathrm{BP}}$, $A_{\mathrm{RP}}$) and compared the GES stars to the parent sample. The GES stars do not present anomalous extinction values. We also checked if the absolute geometric and spectroscopic $K$-band magnitudes, M$_{\mathrm{K, geo}}$ and M$_{\mathrm{K, spec}}$ respectively, for all stars agreed to ensure no unresolved binaries were infiltrating our GES sample and found this to be the case ($|$M$_{K\mathrm{, geo}}-$ M$_{K\mathrm{, spec}}|<0.5$). Their \texttt{ipd\_frac\_multi\_peak} and \texttt{ipd\_frac\_odd\_win} values from \textsl{Gaia} are also 0, which is an indicator for no unresolved binaries.

\section{Results}
\label{sec_results}

\subsection{Ages and Element Abundances}

\begin{figure*}
    \centering
    \includegraphics[width=1\textwidth]{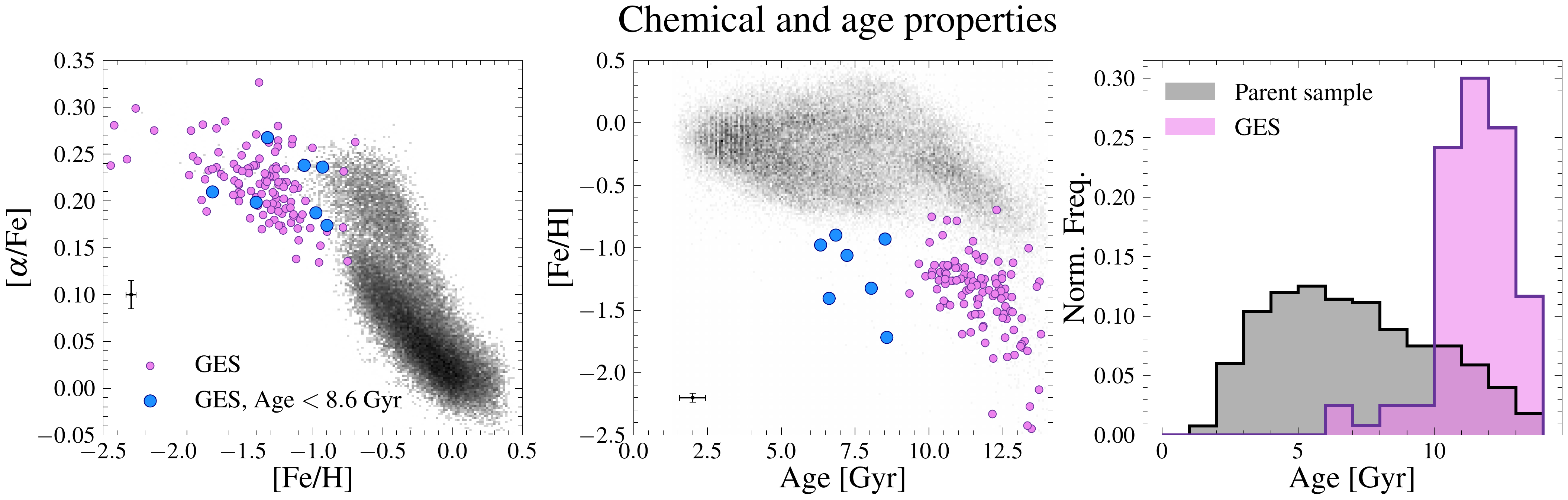}
    \caption{Element abundances and ages for GES stars~(pink) and the younger GES stars with $\mathrm{Age<8.6}$~Gyr~(blue). \textit{Left}: The [$\alpha$/Fe]-[Fe/H] diagram. \textit{Middle}: The [Fe/H]-age plane. \textit{Right}: The age histogram. The GES stars (pink and blue) are distinguishable from the dominant \textit{in situ} disc population~(black) in element abundance-age space. The median age of the GES debris is $=11.6^{+1.14}_{-1.15}$~Gyr, with the range indicating the $16^{\rm th}$ and $84^{\rm th}$ percentiles. The age distribution of GES stars is skewed towards younger ages, reaching $\sim6$--$9$ Gyrs.}
    \label{fig:ages-abun}
\end{figure*}

Fig.~\ref{fig:ages-abun} shows the sample of GES stars~(pink) in the [$\alpha$/Fe]-[Fe/H] plane~(left) and the [Fe/H]-age plane~(middle), as well as a histogram of their age distribution~(right). In all these panels, we also show the parent sample as black. The $\alpha$-Fe compositions of the highly-radial star sample occupy a similar locus to that of the GES, as seen in previous work (\citealp[e.g.,][]{Nissen2010,Hayes2018,Mackereth2019,Monty2020,Horta2021,Horta2023}). The GES stars are metal-poor ([Fe/H] $<-0.9$) and trace an $\alpha$-Fe trajectory that sits below the metal-poor tail of the high-$\alpha$ disc. The age-metallicity distribtion shows how the majority of the GES stars are old (Age $\gtrsim10$ Gyr) and metal-poor, sitting below the main body of the parent sample that is primarily more metal-rich and younger, as expected for disc populations. However, there is a tail of stars in the GES sample that reach younger ages ($6\lesssim\mathrm{Age}\lesssim9$ Gyr, shown in blue), which are also more metal-poor than stars in the parent sample for the same age. The tail of intermediate-age stars in the GES sample is also clearly seen in the right panel of Fig.~\ref{fig:ages-abun}. Here, the median is $11.6^{+1.14}_{-1.15}$~Gyr, with the range indicating the [$16^{\mathrm{th}}$, $84^{\mathrm{th}}$] percentiles of the distribution. However, the distribution is skewed to younger ages. In total, there are seven GES outlier stars with ages below the $5^{\mathrm{th}}$ percentile ($<8.6$~Gyr) of the distribution. 

To ensure that these outliers are statistically significant, we perform two exercises. First, we randomly sample the age of each star from a normal distribution $10^{4}$ times, centered on the XR22 reported age measurement, with a standard deviation equal to the corresponding age error. For each of these $10^{4}$ trials, we compute the $5^{\mathrm{th}}$ percentile of the distribution of the resampled 120 ages, and count the number of stars that fall below this value. For $86\%$ of the trials, we often recover the same seven intermediate-age stars; $11\%$ of the time we recover six stars, and $3\%$ of the time we recover eight stars. Thus, when sampling within the age uncertainties, we still recover the intermediate-age stars. In the second test, we randomly sample with replacement 120 stars from the GES age distribution and calculate the number of times we are able to identify these intermediate-age GES outlier stars. When running this random sampling $10^{4}$ times, the average number of stars that fall below the $5^{\mathrm{th}}$ percentile value of the \emph{original} distribution is $\sim6$--$8$. However, $\sim3\%$ of the time, we only recover $\sim1$--$3$. This result confirms that this tail of younger GES stars is statistically significant.

In the following subsection, we set out to examine the kinematics and orbits of the GES sample and seek to ensure if the younger GES stars follow the same orbital pattern as the main body of the debris.

\subsection{Velocities and Orbits}

\begin{figure*}
    \centering
    \includegraphics[width=1\textwidth]{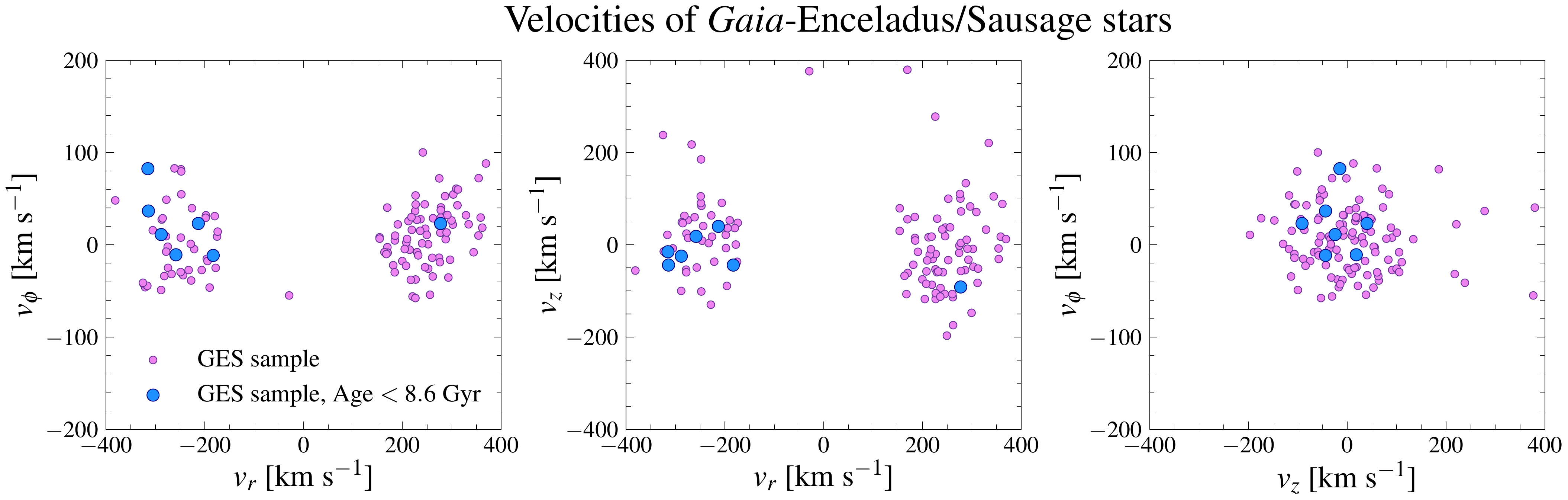}
    \caption{GES stars (pink) in different Galactocentric cylindrical velocity projections. Highlighted as blue points are stars in the GES sample with ${\rm Age} < 8.6$~Gyr.  This cutoff corresponds to the $5^{\mathrm{th}}$ percentile of the age distribution, as shown in the right panel  of Fig.~\ref{fig:ages-abun}. The intermediate-age stars occupy the same locus as the main body of the GES debris in all velocity planes.}
    \label{fig:vel}
\end{figure*}

The GES remnant was initially discovered due to the highly-radial orbits of its debris. One of the kinematic diagrams used to illustrate this point is the $v_r$ vs. $v_\phi$ plane \citep[][]{Belokurov2018}, which relates the radial and azimuthal velocities of stars in a cylindrical Galactocentric coordinate system. The GES stars are shown in this diagram, as well as the $v_R$ vs. $v_z$ and $v_z$ vs. $v_\phi$ planes, in Fig.~\ref{fig:vel}. Here, the full GES sample is shown as pink, and the intermediate-aged stars (i.e., Age $<8.6$ Gyr) are shown in blue. All the GES-selected stars occupy a position in this diagram at either end of the $v_R$ distribution, where one would expect highly-radial orbits. This is to be expected, as the initial selection of the GES sample is tailored to pick out highly-radial orbits, and thus these stars will not present a high signature of rotation (i.e., low-$v_\phi$). Additionally, the sample presents low vertical velocities, $v_z$, implying that the majority of the kinetic energy of these stars is in the radial direction. 

\begin{figure*}
    \centering
    \includegraphics[width=0.9\textwidth]{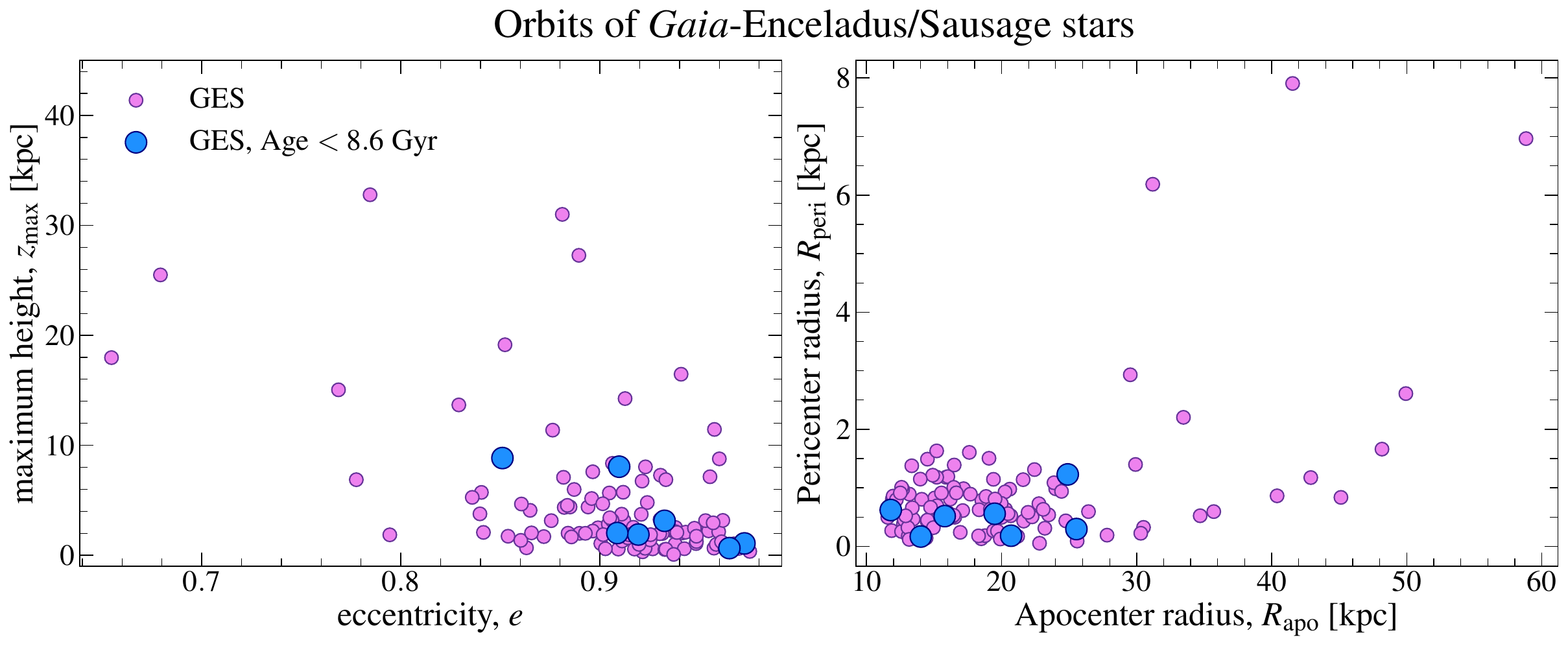}
    \caption{Same as Fig.~\ref{fig:vel}, but now showing orbital properties of stars. In detail, \textit{left} shows the maximum orbital height, $z_{\mathrm{max}}$, as a function of orbital eccentricity, $e$; \textit{right} shows apocenter versus pericenter distance. Both the main body of the GES debris, as well as the intermediate-age stars, present orbits that are not confined to the Galactic plane (i.e., $z_{\mathrm{max}}\gtrsim1$ kpc) and reach large apocenter radii ($R_{\mathrm{apo}}\sim10$--$30$ kpc).}
    \label{fig:ecc}
\end{figure*}

Highly-radial orbits can be found in the plane of the Galaxy due to interactions with the Galactic bar and resonances in the disc. To ensure that the GES sample is comprised of stars on halo-like orbits and thus not confined to the plane, we examine the maximum orbital vertical height, $z_{\mathrm{max}}$, of its stars as a function of eccentricity in the left panel of Fig.~\ref{fig:ecc}.  All the stars in the sample have $z_{\mathrm{max}}\gtrsim1$~kpc and are thus not likely part of the Milky Way disc. They also present high eccentricity ($e>0.8$), by definition. Moreover, their apocentric distances are large ($R_{\mathrm{apo}}>10$ kpc, see right panel of Fig.~\ref{fig:ecc}), whilst their pericentric distances are small ($R_{\mathrm{peri}}<2$ kpc), indicating that these are likely halo stars on eccentric orbits, consistent with the expectation for the GES debris. 

Lastly, in addition to the kinematics and orbits, we inspected the present-day position of these stars and their distance with respect to the position of the Sun. All the GES stars, including the intermediate-age ones, are distributed uniformly within a (present-day) radius of 3~kpc from the Sun.

\section{Discussion}
\label{sec_discussion}


This work demonstrates that the age distribution of the GES stars has a median of $\sim 11.6$~Gyr, with $16^{\rm th}$ and $84^{\rm th}$ percentiles of $\sim 10.5$ and 12.7~Gyr, respectively (Fig.~\ref{fig:ages-abun}).  This corroborates the findings from previous work attempting to age date the GES debris either using CMD fitting (\citealp[e.g.,][]{Helmi2018,Gallart2019,Bonaca2020,Feuillet2021}) or asteroseismology \citep[e.g.,][]{Montalban2021}, which suggests an age for the GES debris that is ancient ($\gtrsim9$--$10$ Gyr ago). However, we have also identified a tail of intermediate-age stars (${\rm Age} \sim6$--$9$~Gyr) that sit within the first $5^{\mathrm{th}}$ percentile of the distribution.

In a recent study \citep{Bonaca2020}, the age distribution of an ``accreted halo'' sample (that is largely similar to the GES sample) was examined using MSTO stars. The authors found a similar profile to the one presented here, whereby the peak of the distribution is old (${\rm Age} > 10$--$11$~Gyr), but skewed to younger ages. This was also seen in \citet{Feuillet2021} and \citet{Grunblatt2021} using red giants. The raw distribution of ages between the MSTO/red giant stars from these studies and the subgiants from this work are in qualitative agreement and suggest a tail of younger ages in the GES debris.

Here, we discuss the implications of the sample of intermediate-age stars in GES and attempt to contextualise the results by discussing a number of possible scenarios that could explain the data.

\subsection{Robustness of the age estimates}

We begin by performing a series of tests to answer two questions: 1) how robust are the age determinations for the intermediate-age GES stars?; 2) what observable is the culprit for a younger age determination when performing XR22's isochrone fitting procedure?

\begin{figure*}
    \centering
    \includegraphics[width=0.9\textwidth]{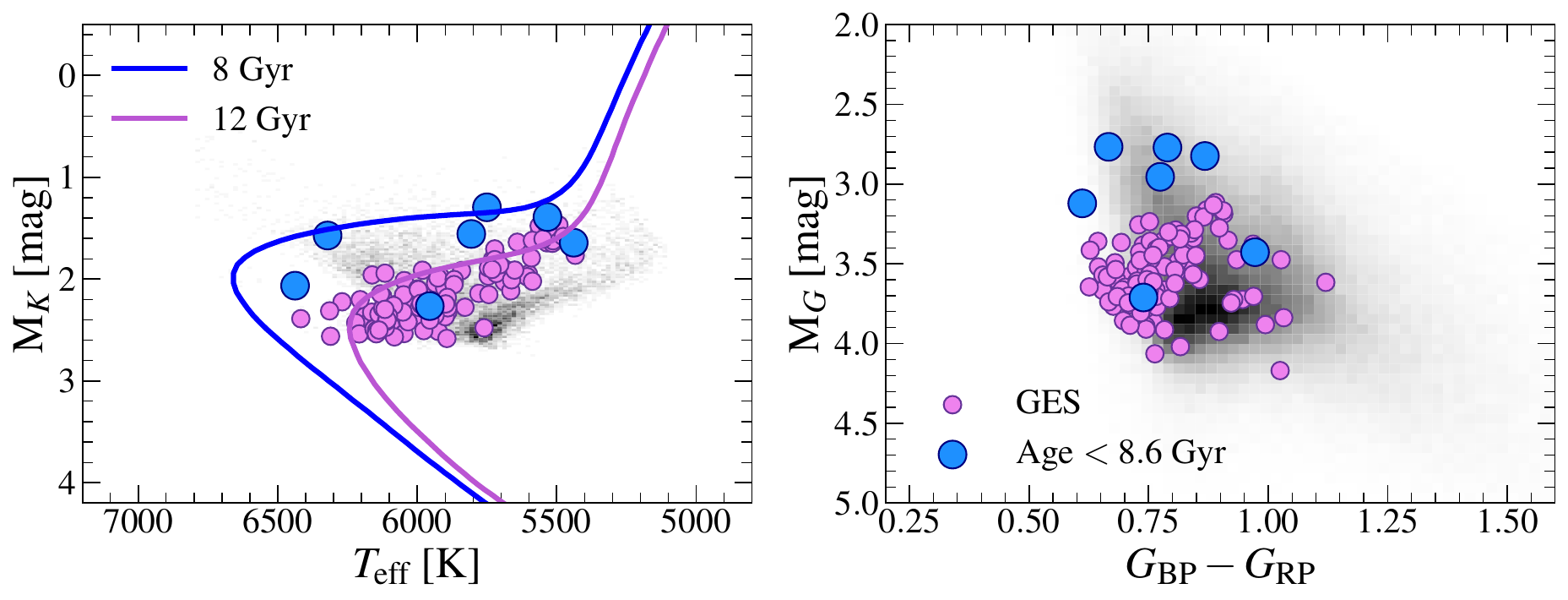}
    \caption{\textit{Left:} $K$-band absolute magnitude, M$_{K}$ (computed as in equation 2 from XR22), as a function of effective temperature, $T_{\mathrm{eff}}$, for the main sample (black), the GES sample (pink), and the intermediate-age GES stars (blue). Also shown on this plot are the 8 and 12 Gyr Yonsei-Yale \citep{Demarque2004} isochrone tracks for a star of [Fe/H] $=-1.5$ and [$\alpha$/Fe] $=0.3$. The average uncertainties in M$_{K}$ and $T_{\mathrm{eff}}$ are $\sim0.04$ mag and $\sim12$ K, respectively. Moreover, the uncertainty in the parallax for all these stars is below $\sim0.04$ arc seconds. \textit{Right:} \textsl{Gaia} color-magnitude diagram (with apparent magnitude, $G$). These figures show how the M$_{K}$ value is higher for the intermediate-age GES stars when compared to the main body of the debris, leading to a younger age determination when performing the isochrone fitting. On the contrary, the \textsl{Gaia} magnitudes and colors appear all similar for both the intermediate-age stars and the main body of the GES debris. }
    \label{fig:checks-age}
\end{figure*}

We set out to compare all the observable properties that are used in XR22's isochrone fitting (namely, $T_{\mathrm{eff}}$, M$_K$, [Fe/H], [$\alpha$/Fe], $G, G_{\mathrm{BP}}, G_{\mathrm{RP}}, J, H, K$) of the intermediate-age stars with the main body of the GES sample. In Fig~\ref{fig:kiel}, we show how the effective temperature, $T_{\mathrm{eff}}$, for the intermediate-age stars overlaps with the main body of the GES debris. This is also the case when examining the [Fe/H] and [$\alpha$/Fe] abundances (Fig~\ref{fig:ages-abun}) and the \textsl{Gaia} and \textsl{2MASS} magnitudes (right panel of Fig~\ref{fig:checks-age}; figures for $J,H,K$ not shown). 

In the left panel of Fig~\ref{fig:checks-age}, we also show the absolute $K$-band magnitude, M$_{K}$, as a function of $T_{\mathrm{eff}}$ for the GES stars (in a similar fashion to Fig. 1 from XR22). Here, we also overlay two Yonsei-Yale isochrone tracks \citep[][]{Demarque2004}, corresponding to stars with [Fe/H] $=-1.5$ and [$\alpha$/Fe] $=0.3$ (approximately the mean of the GES sample). The two tracks correspond to either an age of 8 or 12 Gyr. The intermediate-age GES stars appear brighter, at fixed effective temperature, than the main body of the GES debris, as would be expected for younger populations. Thus, M$_{K}$ is likely the observable that is swaying the age inference for these stars to younger ages when compared to the main body of the GES debris. Conversely, all the other observable properties used to estimate the stellar ages are the same for the intermediate-age GES and the main body of the debris.

\subsection{Binaries and mass transfer leading to incorrect isochrone ages?}

Inaccurate age inferences due to mass transfer from, or coalescence with, a binary stellar companion could explain the intermediate-age stars found in the GES sample. In such a scenario, the resulting primary star would appear brighter than its initial conditions and by default, have a younger inferred age from  isochrone fitting. Peculiar element abundances serve as a key diagnostic to determine if a star has gone through mass transfer and/or has merged with an Asymptotic Giant Branch~(AGB) star companion. To check for this possibility, we use the \textsl{LAMOST} spectra for the seven intermediate-age GES stars, as well as 35 nearest-neighbour stars in stellar parameters\footnote{A  nearest neighbour is defined as a star with near-identical $T_{\mathrm{eff}}$, $\log(g)$, [Fe/H], [$\alpha$/Fe], but which is  old (${\rm Age} >9$~Gyr)---see Appendix~\ref{app_spectra} for more details.} (five for each intermediate-age star) and compare particular Barium (Ba) lines \citep[][]{Guo2023}.  If a low-mass AGB companion enriches the primary with Barium \citep[e.g.,][]{Boffin1988,Han1995} via mass transfer, then the intermediate-age star should show anomalously strong Ba lines compared to the nearest-neighbour reference spectra.  

Fig.~\ref{fig_spec} shows the \textsl{LAMOST} spectra for one of the seven intermediate-age GES stars, around the wavelength of the Ba lines ($\lambda=[4554, 4934, 6495]~\AA$), compared to the five nearest-neighbours. Similar results for the other six intermediate-age GES stars are provided in Appendix~\ref{app_spectra}. None of the seven intermediate-age GES stars have a Ba absorption line that is stronger than its nearest neighbours. Therefore, they likely do not have higher Ba abundances than the rest of the population and are likely not a product of a mass transfer with a low-mass AGB companion. Moreover, for two of these seven stars, there are [Ba/Fe] abundances determined in \textsl{LAMOST}; the abundances of these two stars are not enhanced and are similar to their nearest-neighbour counterparts.

Similarly, unusual features in the spectra and/or double spectral lines can be a sign of stellar binaries. To test if these seven intermediate-age stars present an unusual/double-line feature, we set out to examine a prominent spectral line in the \textsl{LAMOST} spectra (namely, H-$\alpha$). Fig.~\ref{halpha} shows the H-$\alpha$ line of one of the seven stars (the other six are in Appendix~\ref{app_halpha}) in black. For comparison, we also show in grey the spectra for five nearest-neighbour stars (similar to Fig~\ref{fig_spec}). We find no evidence for unusual or double lines in these stars. This supports the hypothesis of these stars not being in binaries.

\begin{figure}
    \centering
    \includegraphics[width=0.9\columnwidth]{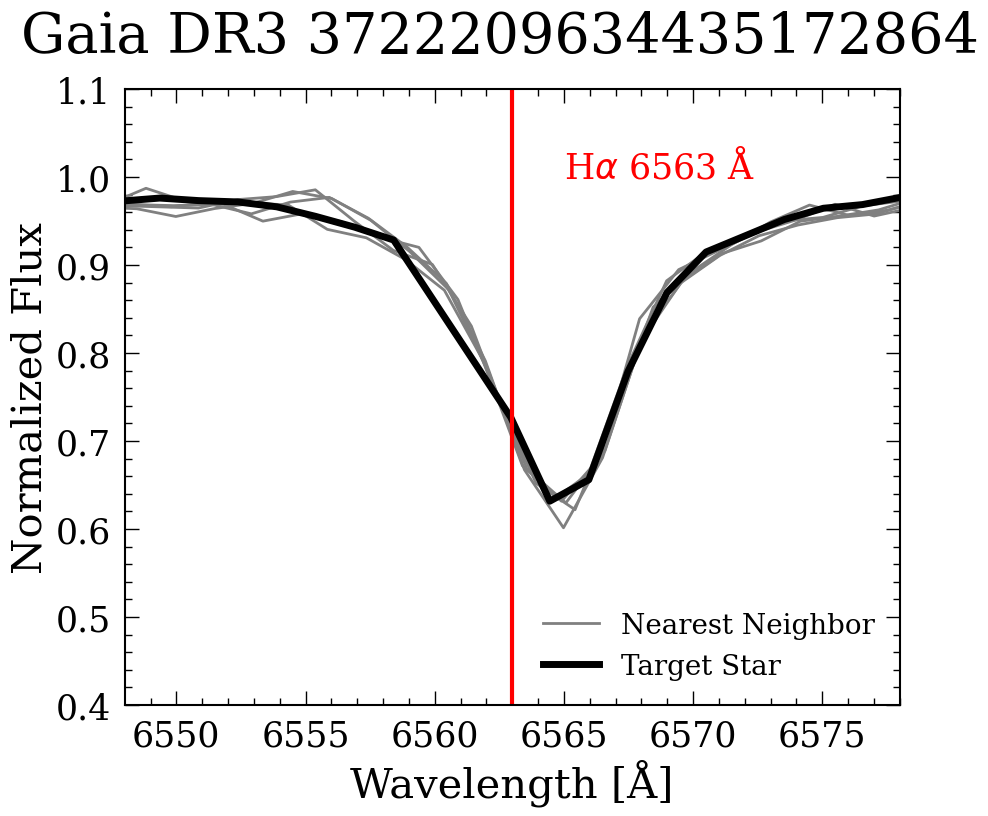}
    \caption{Spectra of one of the seven intermediate-age GES stars (black) around the wavelength region for the H-$\alpha$ line (red) in the \textsl{LAMOST} spectrum. For comparison, we also show the spectra of five nearest-neighbour stars in the catalogue as gray lines (see Appendix~\ref{app_halpha} for details). The seven intermediate-age GES stars do not present any unusual or double lines. Thus, these stars are likely not in a binary. }
    \label{halpha}
\end{figure}

While we find no evidence for mass transfer from a low-mass AGB companion, it could still be that the intermediate-age GES stars experienced mass transfer from a higher-mass AGB or Red Giant Branch~(RGB) and are masquerading as younger stars  (for the same reasons stated above), that cannot be spotted with peculiar element abundances. Follow-up investigation of the apparent young alpha-rich population that comprises around $\sim5$--$10\%$ of the high-alpha disk (\citealp[][]{Chiappini2015,Martig2015}) has demonstrated that a subset of these stars are likely merger or mass-transfer products (\citealp[e.g.,][]{Jofre2016,Hekker2019,yu2024}). This young alpha-rich fraction is aligned with the numbers of intermediate-mass stars in the GES sample. 

In addition to peculiar element abundances, one can also test for binarity looking at line-of-sight-velocity measurements for the same star from different surveys/epocs. If a star is in a binary, its line-of-sight-velocity measurement may show possible shifts in value depending on when the star is observed. To test this, we obtained line-of-sight-velocity measurements from \textsl{Gaia}~DR2, \textsl{APOGEE}~DR17, and \textsl{GALAH}~DR3 for overlapping stars with the \textsl{LAMOST}~DR7 subgiants and compared their measurements. For all 28,863 stars in common (and 27 GES stars\footnote{None of the seven intermediate-age GES stars are contained in this sample of 27 GES stars.}), the line-of-sight-velocity measurements are in good agreement and there are no discrepancies/outliers. While this is a very restricted test of binarity, without intra-survey time-domain data, it is a simple approach available to exclude any stars that do have line-of-sight-velocity differences.

Testing for a white-dwarf companion with the current data is extremely limited and a wider parameter space of binary architectures may be explored with high-resolution time-domain spectroscopy to obtain very precise line-of-sight-velocity measurements. Such data will help establish whether the intermediate-age GES stars are a population that has experienced mass transfer from a  companion and are intrinsically old.

\begin{figure*}
    \centering
    \includegraphics[width=1\textwidth]{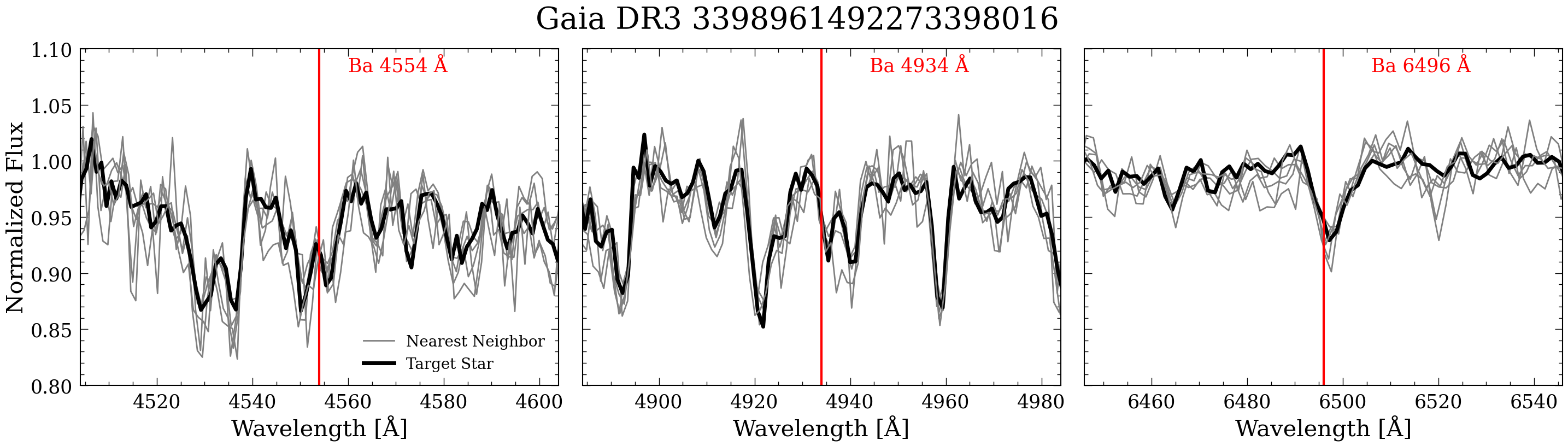}
    \caption{Spectra of one of the seven intermediate-age GES stars (black) around the wavelength region for barium~(Ba) lines (red) in the \textsl{LAMOST} spectrum. For comparison, we also show the spectra of five nearest-neighbour stars in the catalogue as gray lines (see Appendix~\ref{app_spectra} for details). The seven intermediate-age GES stars do not present any strong Ba absorption lines. Thus, these stars are likely not the result of mass transfers from a (low-mass) AGB companion. }
    \label{fig_spec}
\end{figure*}

\subsection{Intermediate-aged stars in the GES progenitor?}

If the intermediate-age stars in the GES sample are not due to incorrect isochrone ages, then it is worth considering the implications for the GES progenitor and the Milky Way's merger history more broadly.

One plausible scenario given our results is that the GES system contains stars that are younger than the time in which most studies conclude it was accreted onto the Milky Way. This scenario would naturally result in the chemical-dynamical properties of the younger GES stars matching the distribution of the main body of the GES debris. Indeed, the intermediate-age stars identified here are [Fe/H]-poor and present [$\alpha$/Fe] abundances that follow the distribution of the main body of GES stars (left panel of Fig.~\ref{fig:ages-abun}; \citealp[e.g.,][]{Nissen2010,Mackereth2019,Monty2020,Horta2023}). They also trace a trajectory in the [Fe/H]-age plane that is distinct from \textit{in situ} populations (see middle panel of Fig.~\ref{fig:ages-abun}). Given our selection criteria, the GES stars present high radial velocities and low azimuthal ones (Fig.~\ref{fig:vel}), typical of the GES debris \citep[][]{Belokurov2018, Helmi2018}. All these results suggest that the $6$--$9$ Gyr stars are in fact part of the GES debris. 

If the intermediate-age stars are part of the GES system, they were formed prior to when the progenitor galaxy quenched star formation. Recent theoretical work has shown that large pericentric passages of satellites during their first infall onto larger-mass hosts can induce star formation in the infalling satellite (\citealp[e.g.,][]{Pasetto2011,Rocha2012,Dicintio2021}). This is expected to be the case for the Sagittarius dwarf spheroidal~(Sgr dSph) \citep[e.g.,][]{Siegel2007,Boer2015} and the LMC (\citealp[][]{Nidever2020,Hasselquist2021}). Thus, one feasible explanation for the presence of these younger GES stars is a burst of star formation during infall of the GES progenitor galaxy onto the Milky Way. However, this scenario would also imply that the GES merger event was not short-lived, but instead occurred over a longer timescale. It also likely implies that the GES merger was an accretion episode that endured more than one pericentric passage. While different types of accreted satellites can yield highly-radial orbit debris (\citealp[][]{Amorisco2017,Khoperskov2023,Horta2023_fire}), radialization is most effective for satellites with high initial eccentricity \citep[]{Vasiliev2022} that typically are more short-lived. Interestingly, in a recent study, \citet{Naidu2021_ges} ran a high number of tailored $N$-body simulations and found that the GES debris are most closely matched to a satellite that was accreted on a retrograde orbit and underwent three (close) pericentric passages, which then rapidly radialized. While these results should be taken with caution as these simulations were run in an idealised set-up (i.e., no other mergers and no Milky Way galaxy growth), they highlight that the GES debris could be explained by a merger event with multiple pericentric passages.

Similarly, \citet{Amarante2022} explored tailored and idealised galaxy merger scenarios of GES-like systems onto a Milky Way-mass galaxy, including a gas and star formation prescription. The authors also found that the GES-like mergers undergo two-to-three close pericentric passages before fully radializing. However, during this merger process, the infalling satellites do not form any stars after \texttt{t$_{\mathrm{merge}}$} --- the time in which the satellite fully disrupts and becomes phase-mixed.

Furthermore, it is also plausible that, instead of a star formation burst in the GES being induced by first infall onto the Milky Way, that a burst in star formation could have been provoked by the interaction of the GES progenitor and a separate infalling satellite that accreted at a similar time onto the Galaxy. There have been additional debris/mergers proposed in the literature that present radial orbits, conjectured to have possibly fallen in with the GES (e.g., \textsl{Sequoia}: \citealp[][]{Barba2019,Myeong2019}) that could satisfy this scenario. 

Another possible scenario could be that the GES merger event is not as ancient as once thought (namely, $\gtrsim10$ Gyr ago) and in fact accreted onto the Milky Way at a later time, approximately $6$--$9$ Gyr ago. If the gas pertaining to the GES galaxy was stripped during accretion onto the Milky Way, the GES system would have ceased star formation, and thus would not have formed any intermediate-age stars after the merger event. Naturally, the presence of stars with ages below the infall time (suggested to be $\gtrsim10$ Gyr) would rule out this scenario. Moreover, recent dynamical work has suggested that the GES debris can be explained from a younger merger event (\citealp[e.g.,][]{Donlon2022,Donlon2023}). Thus, a scenario where the GES progenitor(s) merged with the Milky Way approximately $\sim6$--$9$ Gyr ago could explain the observed intermediate-age stars we see in this work, assuming that the majority of the stars in the GES formed well before infall onto the Milky Way. Furthermore, if the mass ratios estimated for the GES are consistent with it being a major accretion event, a younger merger scenario would also imply that the mass of the GES system is greater than current estimates (\citealp[e.g.,][]{Mackereth2020,Lane2023}).  Conversely, if the GES merger happened more recently than previously thought and the current mass estimates are correct, it would imply that the mass ratio of this merger was smaller.

Lastly, we note that to describe the younger stars in their ``accreted halo'' sample, \cite{Bonaca2020} used a simple one-parameter model with constant star formation rate that shuts off at some end time, $t_{\rm end}$.  Fitting to the data, they found that $t_{\rm end} \gtrsim10$~Gyr ago. While a one-parameter model is a good starting point to understand the star formation history of the GES system, we argue that it may be worthwhile to examine this data with more complex Galactic chemical evolution models as well (see for example \citealp[][]{Hasselquist2021,Johnson2023}).

\subsection{Another sausage in the GES debris?}

Intermediate-age stars with GES-like orbits could be explained by additional younger debris from a separate merger event. Expectations from cosmological simulations suggest that accreted satellites of different mass and infall time can manifest radial orbits (\citealp[e.g.,][]{Amorisco2017,Fattahi2020,Horta2023_fire,Khoperskov2023,Orkney2023, Rey2023}). However, radialization is most efficient in cases of high satellite mass, not very steep host density profiles, and high initial eccentricity \citep[][]{Vasiliev2022}. While it is entirely plausible that additional debris could be infiltrating the GES sample and could help explain the intermediate-age stars we observe (\citealp[e.g.,][]{Donlon2022,Donlon2023}), the sequence in the [$\alpha$/Fe]-[Fe/H] plane and the [Fe/H]-age relation would suggest that the additional merger would have had a similar star formation history to the GES and thus would be of similar mass \citep[e.g.,][]{Cunningham2022,Horta2023_fire,Khoperskov2023}.
If this is the case, it would also imply that a considerable fraction of the old stars in the GES are also part of this additional debris, bringing the overall mass of the GES down (\citealt{Lane2023}). 

\subsection{Stripped stars from the Sagittarius dSph galaxy?}

Another possible scenario is that the intermediate-age stars on highly-radial orbits are the result of stars that were stripped from the infalling Sgr dSph galaxy, which have somehow dynamically evolved to have the same radial orbits as the GES debris. This scenario would (qualitatively) satisfy the element abundances observed (\citealp[e.g.,][]{Hasselquist2021,Horta2023}). However, we rule out this possibility based on the fact that the stars in the GES sample, including the the intermediate-age stars, all show present-day positions within $\sim3$ kpc from the Sun.

\subsection{Heated disc stars?}

Intermediate-age stars on highly-radial orbits could be the result of heated disc populations due to interactions with satellite galaxies. This has been shown to be the case in the Milky Way for the high-$\alpha$ (\citealp[][]{Bonaca2017,Belokurov2020}) and low-$\alpha$ discs (\citealp[][]{Laporte2020,Laporte2022,Das2024}) and could be the result of the GES merger or the Sgr dSph one. This scenario is less likely as the intermediate-age and highly-radial stars occupy a sequence in the [$\alpha$/Fe]-[Fe/H] and [Fe/H]-age planes that does not overlap with either of the high/low-$\alpha$ discs, but does overlap with the main body of the GES debris. The abundance and age properties of these stars suggest they are (accreted) stellar halo populations.

\section{Conclusions}
\label{sec_conclusions}

In summary, using the subgiant catalogue of ages and element abundances provided by XR22 and orbits determined in this work, we explored the age distribution of stars in the Galactic stellar halo on highly-radial orbits. We selected stars in this sample that are likely to form part of the debris from the GES accretion event and found that the majority are old, with median age of $11.6^{+1.14}_{-1.15}$~Gyr, where the range indicates the $16^{\mathrm{th}}$ and $84^{\mathrm{th}}$ percentiles. We also identified a tail of younger outlier stars, with ages down to $\sim6$--$9$ Gyr, below the $5^{\mathrm{th}}$ percentile of the distribution. The [$\alpha$/Fe]-[Fe/H] element abundances, [Fe/H]-age relation, and orbital properties of these $\sim6$--$9$~Gyr stars overlap with the main body of the GES debris. 

One potential explanation for these intermediate-age stars is that they are the product of mass transfer and/or stellar mergers, masquerading as young stars when they are intrinsically old.  While initial tests do not find evidence for this, we cannot rule out this scenario completely with current data.  If the ages of these outliers are indeed well-measured and these stars were formed in, and belong to, the debris of the GES accretion event, it could imply that either: $i$) the GES event is either not as ancient as previously conjectured; $ii$) the GES merger formed stars during infall onto the Milky Way and/or due to interactions with a satellite of its own. The last possible scenario we conjecture is that the local stellar halo is comprised by the debris of two satellites (the GES and a companion debris with younger stars). 

While our results corroborate previous stellar halo age studies \citep[e.g.,][]{Bonaca2020, Feuillet2021, Grunblatt2021}, we emphasize that we have only identified seven intermediate-age GES stars. Thus, these findings should be revisited with larger sample statistics and ideally with subgiant and/or MSTO stars, from which more reliable ages can be determined. Moreover, calibrations of [C/N]-based ages for RGB stars in the metal-poor regime will aid in discerning the age distribution of stellar halo populations. It would be interesting to follow up this study with data from the \textsl{WEAVE} \citep[][]{Dalton2014} and \textsl{Milky Way Mapper}/SDSS-V \citep[][]{Kollmeier2017} programmes that deliver the required spectro-photometric information to obtain stellar ages for GES stars. With these data, it will likely be possible to constrain the origin of these intriguing intermediate-age, metal-poor, and highly-radial stars in the Milky Way's stellar halo.  

\section*{Acknowledgements}
DH thanks Sue, Alex, and Debra. He also thanks Ted A. von Hippel, Carrie Filion, Conny Aerts, Ricardo P. Schiavon, and the Nearby Universe group at the Flatiron Instittue for helpful discussions. The Flatiron Institute is funded by the Simons Foundation.  ML is supported by the Department of Energy~(DOE) under Award Number DE-SC0007968 as well as the Simons Investigator in Physics Award.  

\section*{Data Availability}
All data used in this work was compiled by \citet{Xiang2022} and is publicly available at
\url{https://keeper.mpdl.mpg.de/d/019ec71212934847bfed/}. The orbital quantities used in this work for the parent sample of 11,336 stars can be found at \url{https://www.dropbox.com/scl/fi/t9z8nhtur9591mvfjo91k/subgiant-paper-orbits-gala22.fits?rlkey=m2ppam8iyn67tc4au6c2f65uz&dl=0}.

\software{
    matplotlib \citep{Hunter:2007},
    numpy \citep{numpy},
    \textsl{Gala} \citep{gala2017},
    \textsl{AGAMA} \citep{agama2019}
}

\bibliography{refs}
\clearpage
\appendix
\setcounter{equation}{0}
\setcounter{figure}{0} 
\setcounter{table}{0}
\renewcommand{\theequation}{A\arabic{equation}}
\renewcommand{\thefigure}{A\arabic{figure}}
\renewcommand{\thetable}{A\arabic{table}}

\section{Testing for Evidence of Mass Transfer from AGB Stars}
\label{app_spectra}

\begin{figure*}[h!]
    \centering
    \includegraphics[width=1\textwidth]{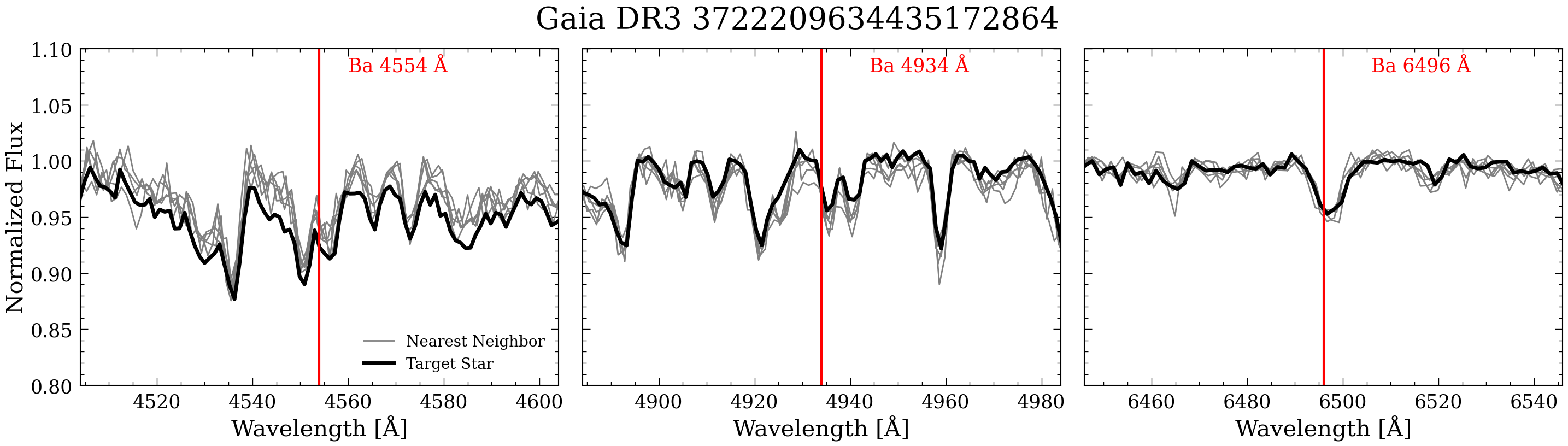}
    \caption{Fig.~\ref{fig_spec} continued.}
    \label{fig_spec_2}
\end{figure*}

\begin{figure*}[h!]
    \centering
    \includegraphics[width=1\textwidth]{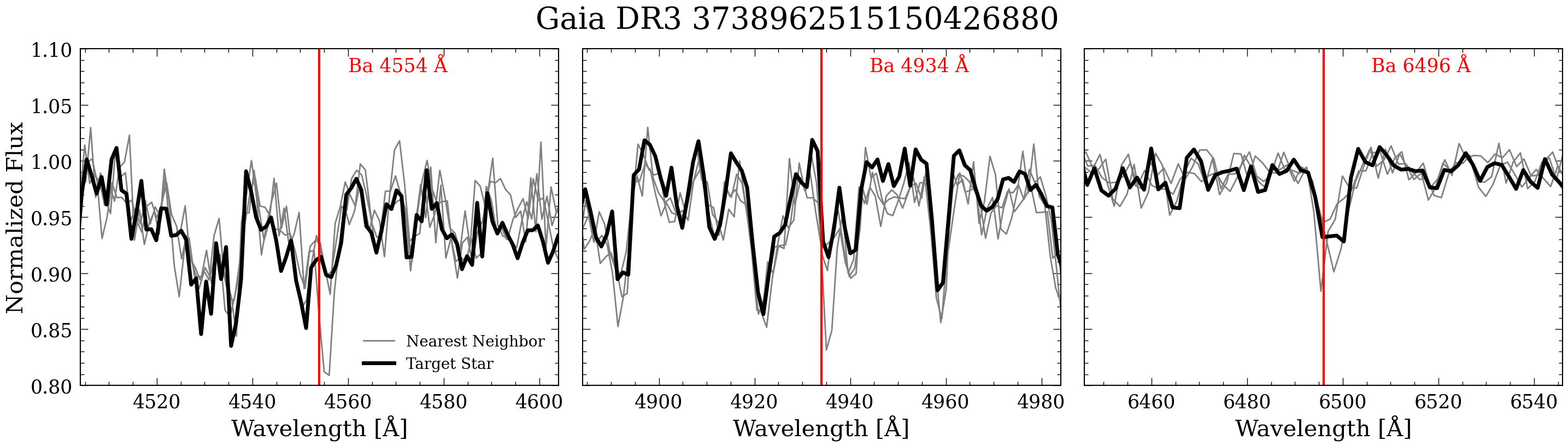}
    \caption{Fig.~\ref{fig_spec} continued.}
    \label{fig_spec_3}
\end{figure*}

\begin{figure*}[h!]
    \centering
    \includegraphics[width=1\textwidth]{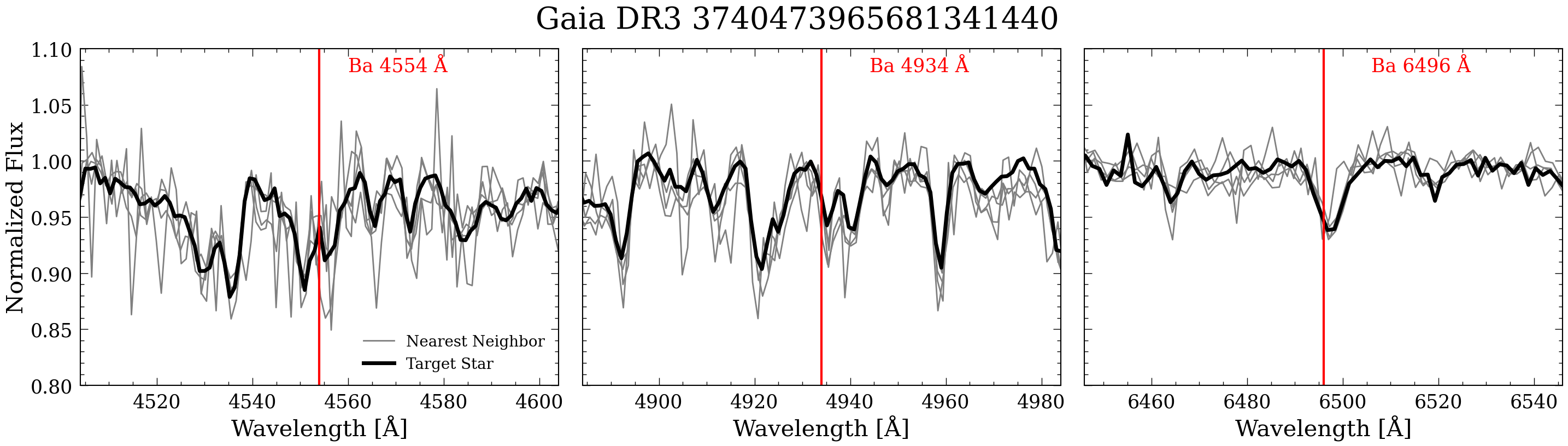}
    \caption{Fig.~\ref{fig_spec} continued.}
    \label{fig_spec_4}
\end{figure*}

\begin{figure*}[h!]
    \centering
    \includegraphics[width=1\textwidth]{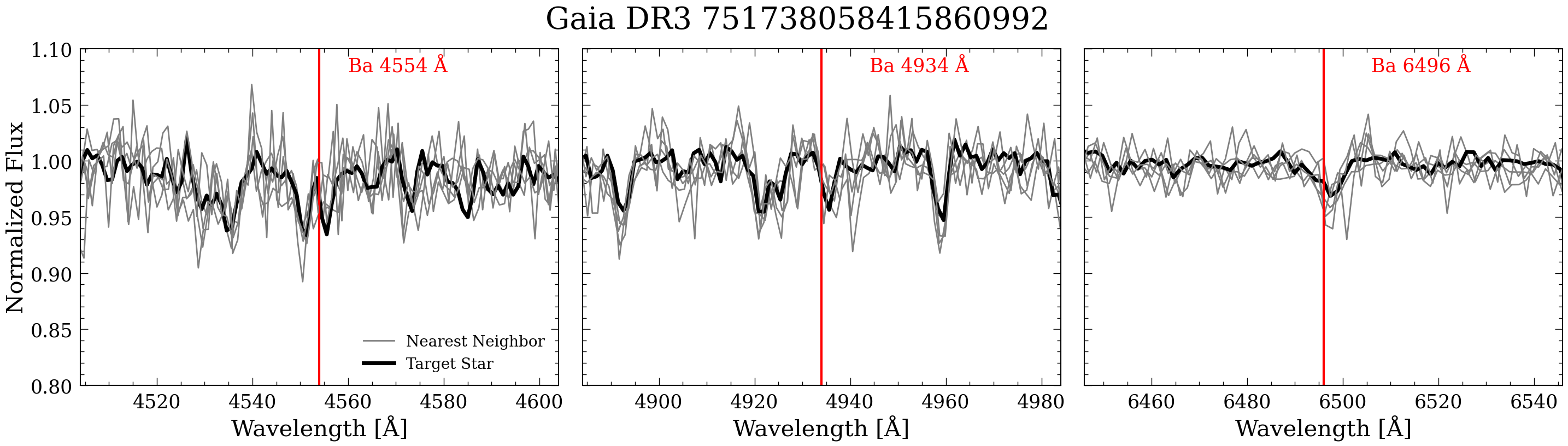}
    \caption{Fig.~\ref{fig_spec} continued.}
    \label{fig_spec_5}
\end{figure*}

\begin{figure*}[h!]
    \centering
    \includegraphics[width=1\textwidth]{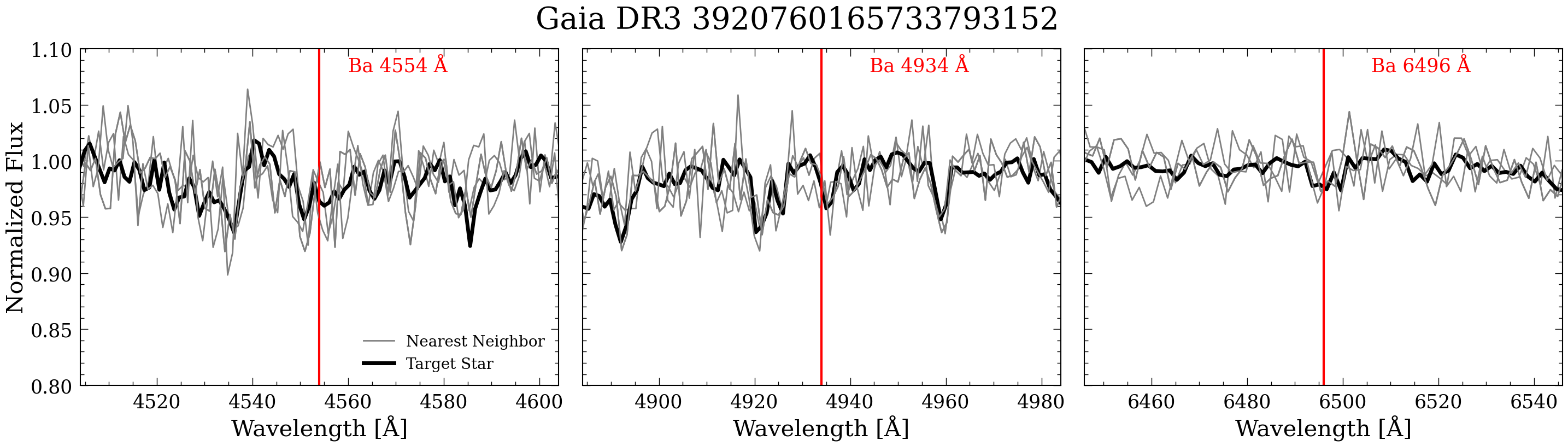}
    \caption{Fig.~\ref{fig_spec} continued.}
    \label{fig_spec_6}
\end{figure*}

\begin{figure*}[h!]
    \centering
    \includegraphics[width=1\textwidth]{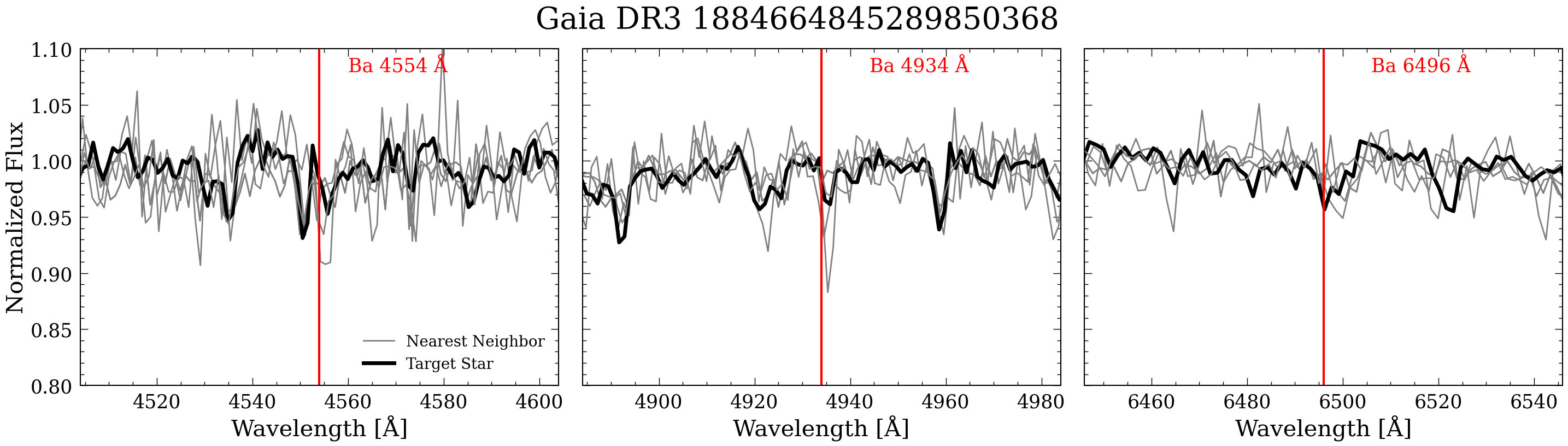}
    \caption{Fig.~\ref{fig_spec} continued.}
    \label{fig_spec_7}
\end{figure*}

Figs.~\ref{fig_spec_2}--\ref{fig_spec_7} are the same as Fig.~\ref{fig_spec}, but for the remaining six intermediate-age GES stars~(black) and their associated nearest neighbors~(gray). The nearest neighbors~(nn) are selected using the following criteria: $|\mathrm{[Fe/H]_{star}} - \mathrm{[Fe/H]_{nn}}| < 0.05$ dex; $|\mathrm{[\alpha/Fe]_{star}} - \mathrm{[\alpha/Fe]_{nn}}| < 0.05$ dex; $|T\mathrm{_{eff, star}} - T\mathrm{_{eff, nn}}| < 50$ K; $|\log(g_{\rm star}) - \log(g_{\rm nn})| < 0.05$ dex; SNR$>50$.
To obtain enough nearest neighbors, the selection criteria are loosened for two stars. 
These two stars and the different selection criteria for their nearest neighbors are: 1) for Gaia DR3 source ID 3920760165733793152, we select stars with $\mathrm{[Fe/H]}$, $\mathrm{[\alpha/Fe]}$, $\log(g)$ within 0.1~dex of the targeted stars, $T\mathrm{_{eff}}$ within 100~K, and SNR $>$10;\footnote{This still yields only four nearest neighbors stars with a median age of 7.5~Gyr.} and 2) for Gaia DR3 source ID 751738058415860992 and Gaia DR3 source ID 1884664845289850368, we select stars with SNR $>30$.

\clearpage

\section{Testing for Evidence of stars in binaries}
\label{app_halpha}

\begin{figure}
    \centering
    \includegraphics[width=0.4\columnwidth]{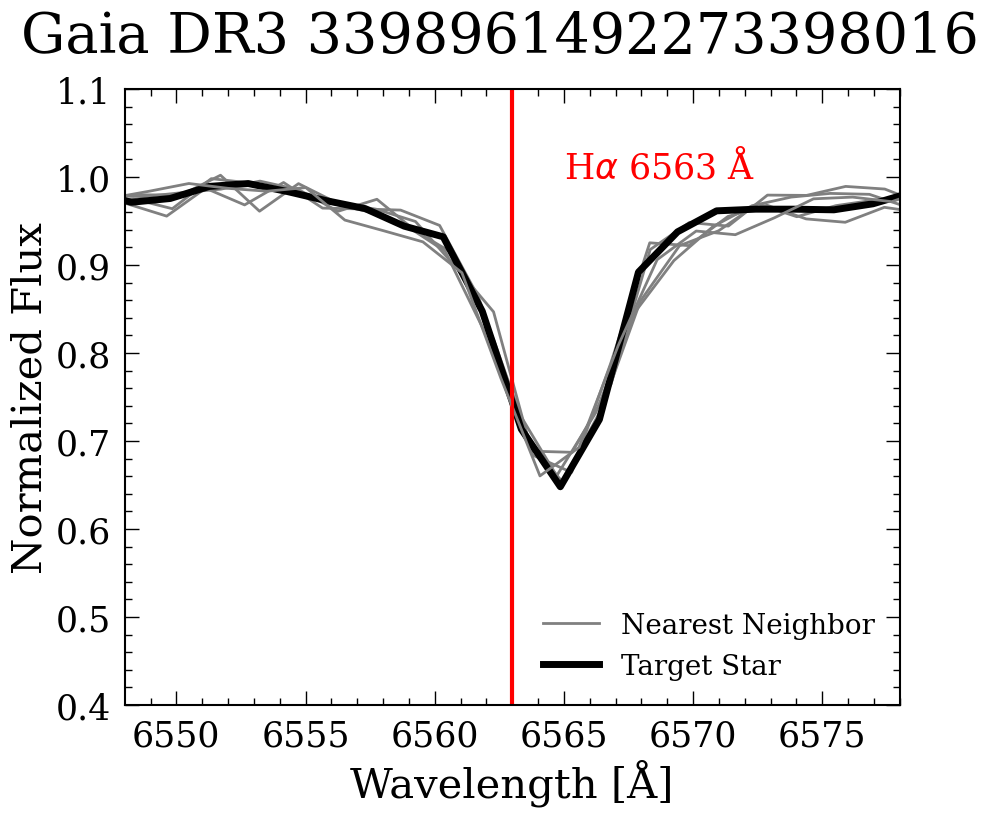}
    \caption{Fig.~\ref{halpha} continued.}
    \label{fig_halpha_2}
\end{figure}

\begin{figure}
    \centering
    \includegraphics[width=0.4\columnwidth]{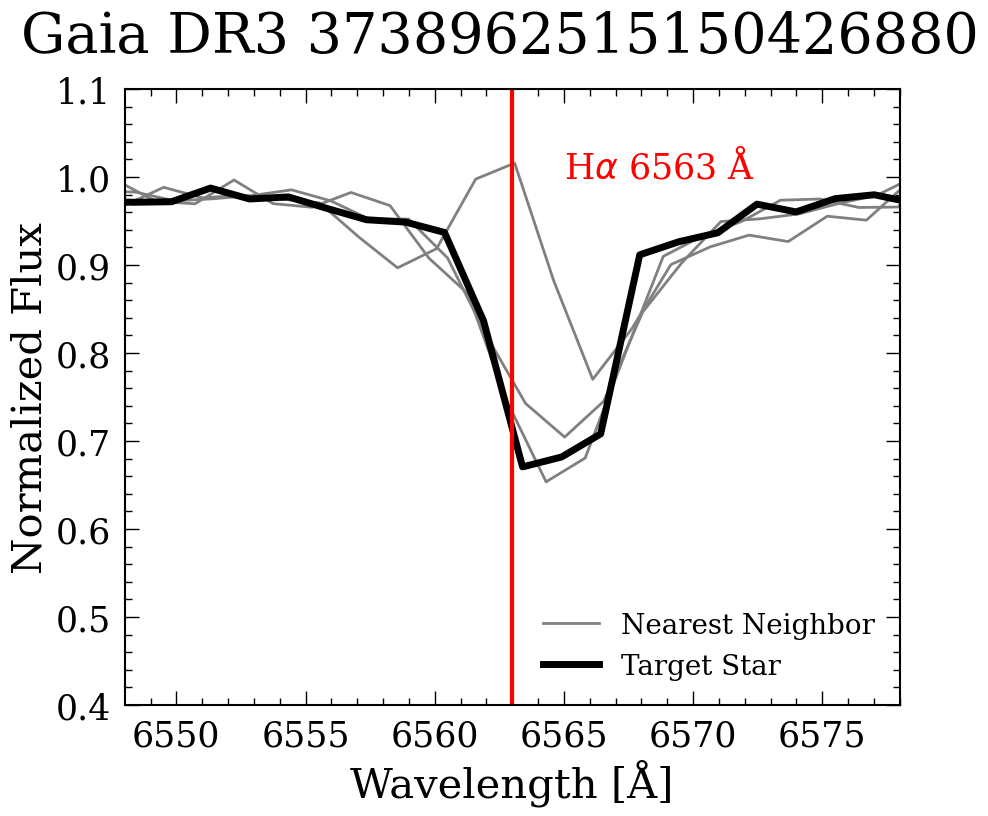}
    \caption{Fig.~\ref{fig_spec} continued.}
    \label{fig_halpha_3}
\end{figure}

\begin{figure}
    \centering
    \includegraphics[width=0.4\columnwidth]{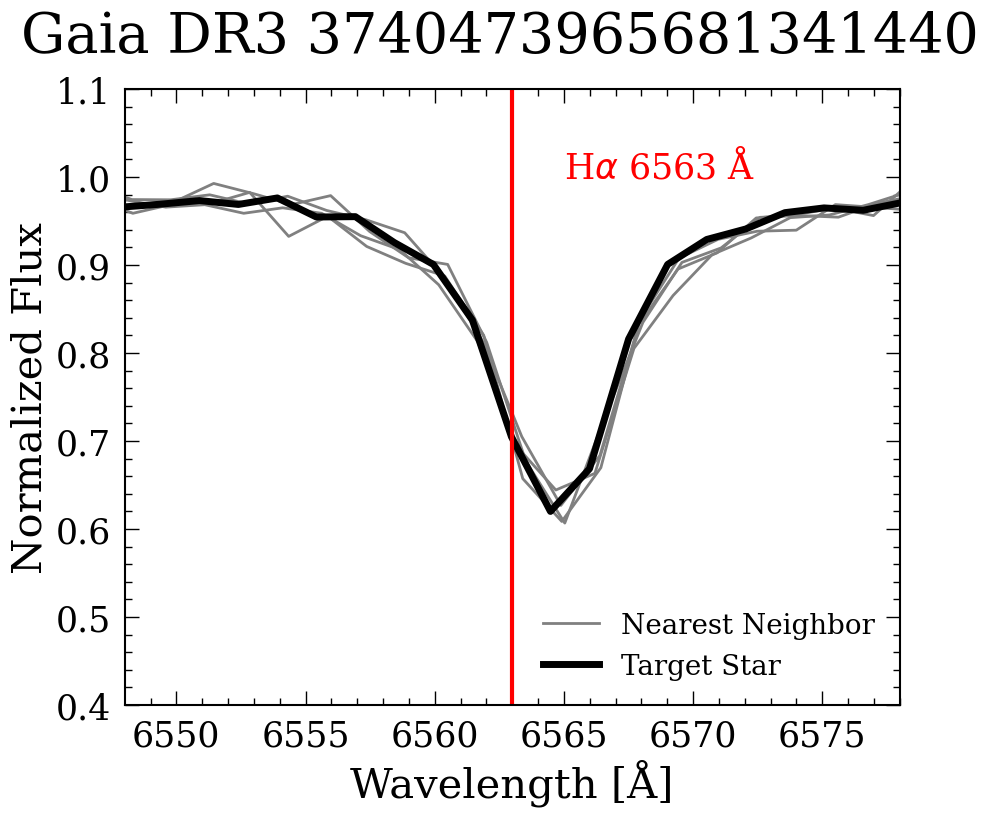}
    \caption{Fig.~\ref{fig_spec} continued.}
    \label{fig_halpha_4}
\end{figure}

\begin{figure}[h!]
    \centering
    \includegraphics[width=0.4\columnwidth]{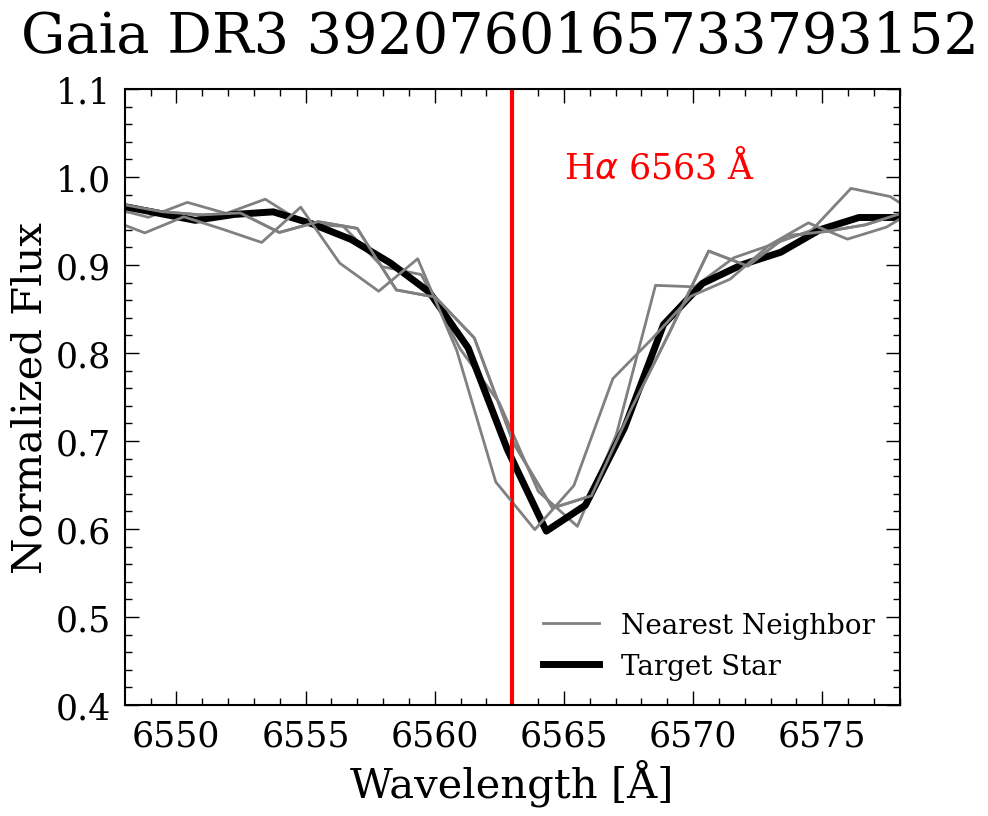}
    \caption{Fig.~\ref{fig_spec} continued.}
    \label{fig_halpha_5}
\end{figure}

\begin{figure}[h!]
    \centering
    \includegraphics[width=0.4\columnwidth]{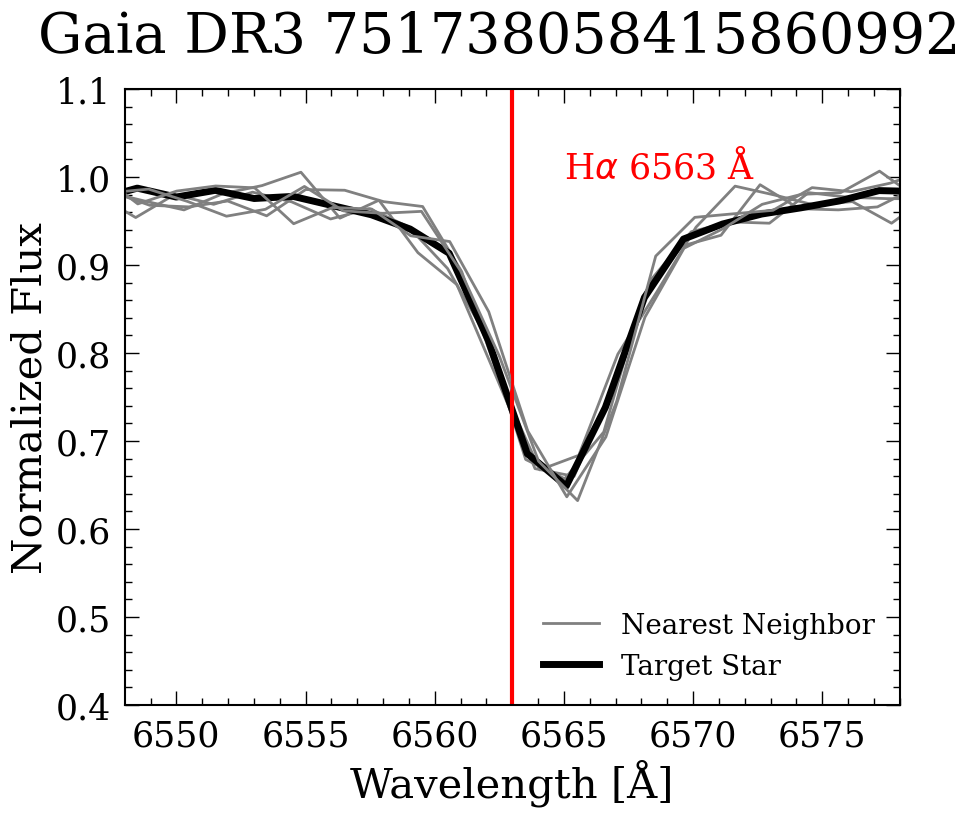}
    \caption{Fig.~\ref{fig_spec} continued.}
    \label{fig_halpha_6}
\end{figure}

\begin{figure}[h!]
    \centering
    \includegraphics[width=0.4\columnwidth]{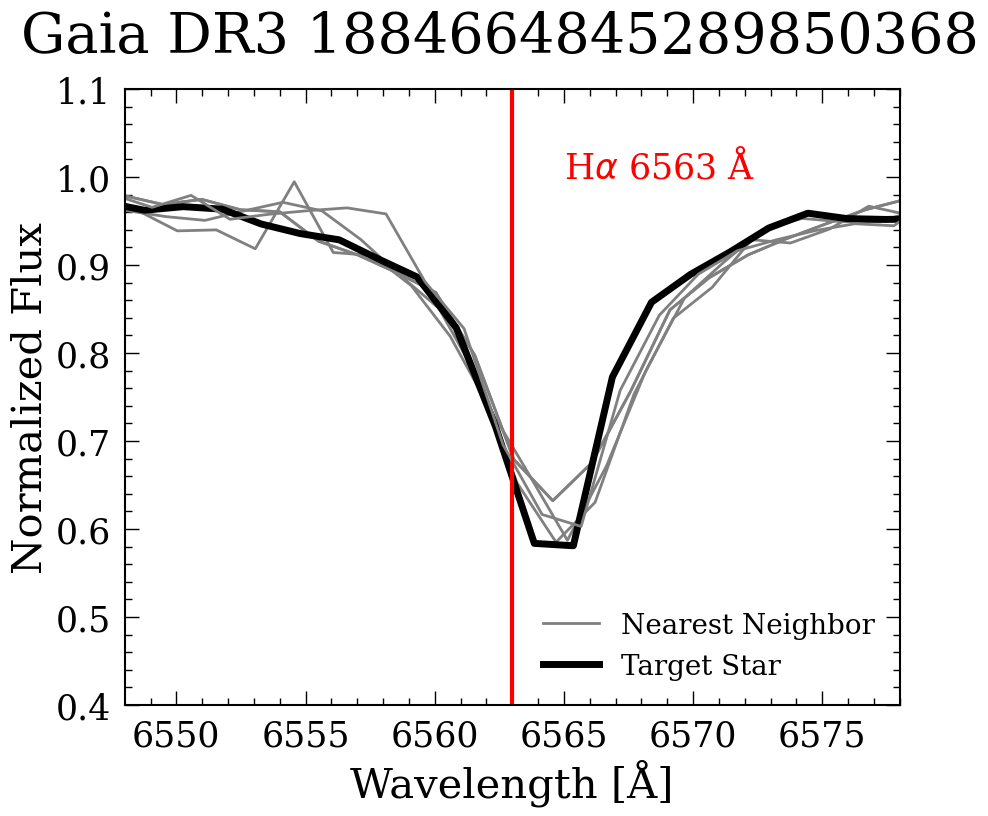}
    \caption{Fig.~\ref{fig_spec} continued.}
    \label{fig_halpha_7}
\end{figure}

\clearpage

\section{Supplementary material: catalogue of orbits}
Table~\ref{tab_data}-\ref{tab_data2} shows the information contained in the table of orbits compiled in this work for the 11,336 stars in our parent sample, which is publicly available at \url{https://www.dropbox.com/scl/fi/t9z8nhtur9591mvfjo91k/subgiant-paper-orbits-gala22.fits?rlkey=m2ppam8iyn67tc4au6c2f65uz&dl=0}.

\begin{table*}
	\centering
	\caption{From left to right, \textsl{Gaia} DR3 ID, \textsl{LAMOST} DR7 ID, orbital actions, angular momentum, orbital energy, pericenter radius, apocenter radius, maximum height, eccentricity, and a boolean mask to select GES stars (options are either ``PARENT'' or ``GES''). Here we show the value for the first three stars. The full table can be downloaded using the link in the main text.}
	\begin{tabular}{lcccccr} 
		\hline
		 \textsl{Gaia} DR3 ID & \textsl{LAMOST} DR7 ID & ($J_{R}, J_{\phi}, J_{z}$)& ($L_{x}, L_{y}, L_{z}$)\\
   & & [kpc km s$^{-1}$] & [kpc km s$^{-1}$]\\
   \hline
    303792868026997248  & 20151001-HD014422N323057V01-02-183   &(52.59, -1969.25, 9.79)  &(0.115, 16.22, -1969.25) \\
    303857739213031296 & 20151001-HD014422N323057B01-05-250 & (54.18, -2205.91, 14.22) & (-17.84, 19.98, -2205.91)\\
    303909240165205632  &20151001-HD014422N323057V01-04-179 & (89.76, -2392.29, 11.78) & (-24.60, 12.85, -2392.29)\\
\hline
	\end{tabular}
 \label{tab_data}
\end{table*}

\begin{table*}
	\centering
	\caption{Table~\ref{tab_data} continued.}
\begin{tabular}{lccccccr}
 \hline
Energy & $R_{\mathrm{peri}}$&$R_{\mathrm{apo}}$& $z_{\mathrm{max}}$& $e$& ``IN$_{-}$GES''\\

[$\times10^{5}$ km$^{2}$ s$^{-2}$]& [kpc] & [kpc] & [kpc] \\
   \hline
-1.23  &7.26   & 10.74   &0.62 &  0.19 & PARENT\\
-1.18 & 8.28 &   12.07 &  0.86 &  0.18 & PARENT\\
-1.13 & 8.68  & 13.85  & 0.85 & 0.23 & PARENT\\
\hline
	\end{tabular}
 \label{tab_data2}
\end{table*}

\end{document}